\documentclass[twocolumn,prl,aps]{revtex4}
\usepackage{graphicx}
\usepackage{dcolumn}
\usepackage{amsmath}

\newcommand{\figurewidth}{8.3cm}
\newcommand{\beq}{\begin{equation}}
\newcommand{\eeq}{\end{equation}}

\begin{document}

\title[Short Title]{
First Principles Calculations of Shock Compressed Fluid Helium}

\author{B. Militzer}
\affiliation{Geophysical Laboratory, Carnegie Institution of Washington,
          5251 Broad Branch Road, NW, Washington, DC 20015} 

\begin{abstract}
The properties of hot dense helium at megabar pressures were studied
with two first-principles computer simulation techniques, path
integral Monte Carlo and density functional molecular dynamics. The
simulations predicted that the compressibility of helium is
substantially increased by electronic excitations that are present in
the hot fluid at thermodynamic equilibrium. A maximum compression
ratio of 5.24(4)-fold the initial density was predicted for 360 GPa
and 150$\,$000 K. This result distinguishes helium from deuterium, for
which simulations predicted a maximum compression ratio of
4.3(1). Hugoniot curves for statically precompressed samples are also
discussed.
\end{abstract}

\date{\today }


\maketitle


There has been considerable controversy in the deuterium equation of
state (EOS) since laser shock wave experiments probed the megabar
pressure regime for the first time and predicted that deuterium is
highly compressible under shock conditions to approximately 6-fold the
initial density~\cite{Si97,Co98}. Such a high compression ratio was
neither reproduced with magnetic compression
experiments~\cite{Kn01,Kn03} nor with explosively driven
shocks~\cite{Belov2002,BoriskovNellis05}. Both sets of later
experiments predicted compression ratios close to 4.3(1), which is in
good agreement with results from first principles computer
simulations~\cite{Le97,MC00,Bonev2004}. When we applied the same
simulation techniques, path integral Monte Carlo (PIMC) and density
functional molecular dynamics (DFT-MD), to hot dense helium, we found
that helium's shock compressibility is substantially increased due to
electronic excitations in the fluid.

In this Letter, we make the prediction that electronic excitations in
helium lead to a maximum shock compression ratio of 5.24(4), while
such excitations in deuterium do not increase the compressibility
ratio beyond 4.3(1). Furthermore, we show that the compression ratio
is reduced when the sample is precompressed statically in a diamond
anvil cell before a shock is launched. Such novel compression
techniques are currently developed and data for dense helium are
forthcoming~\cite{LoubeyreAPS}. The combination of static and dynamic
compression techniques allows the study of materials at much higher
densities, and their application to hydrogen and helium will enable a
direct characterization of a much larger section of the isentrope that
determines the interiors of giant planets.

Present studies of giant planetary interiors~\cite{Gu02} are based on
approximate free energy models that rely on analytical thermodynamic
expressions and are often fit to experimental results if
available. Although these models (for helium
read~\cite{SC95,SJR05,WC05}) are very practical, their predictive
capabilities for shock states and the EOS are limited because
interaction and polarization effects in a dense fluid are very
difficult to study analytically, which underlines the need of first
principles simulations.


Shock wave experiments provide us with direct information for
materials' EOS at high pressure and temperature. When a shock wave
passes through the sample, the thermodynamic state of the material,
characterized by the internal energy, pressure, and volume, changes
from initially ($E_0,P_0,V_0$) to the final values of ($E,P,V)$. The
conservation of mass, momentum, and energy yields the Hugoniot
condition~\cite{Ze66},
\beq
H = (E-E_0) + \frac{1}{2} (P+P_0)(V-V_0) = 0.
\label{hug}
\eeq
The resulting Hugoniot curve is the locus of all final states that can
be reached for different shock velocities. Theoretically, the Hugoniot
curve can be calculated from the EOS, assuming $P_0 \ll P$ and using
the initial molecular volume from the experiment, $V_0=32.4$ {\rm
cm}$^3$/mol ($\rho_0=0.1235\,$g$\,$cm$^{-3}$)~\cite{nellis84}. For
$E_0$ one takes the energy of an isolated helium atom, which must be
consistent with the internal energy $E(V,T)$ derived from a particular
method. For PIMC we use $E_0=-$79.0048~eV per atom because PIMC is
exact for the helium atom. The assumptions for $P_0$ and $E_0$ also
remain sufficiently valid for samples that have been precompressed
statically to 4 times $\rho_0$.

PIMC is a finite-temperature quantum simulation method that we used to
model dense fluid helium as a system of nuclei and electrons that
interact via the Coulomb potential. Both types of particles are
explicitly treated as paths and all correlation effects are included,
which makes PIMC one of the most accurate finite-temperature quantum
simulation methods available. The only noncontrolled approximation
required is the fixed node approximation that is introduced to treat
the fermion sign problem, which arises from the explicit treatment of
electrons. The fermion nodes are taken from a thermal trial density
matrix, for which we extended the variational density matrix
approach~\cite{MP00} to helium.

We complement our PIMC EOS with DFT-MD data, since DFT is much more
efficient at low temperatures because it is in principle a ground
state electronic structure method.  The DFT-MD trajectories were
obtained with Born-Oppenheimer MD where the electrons were assumed to
be in the instantaneous ground state. We used the CPMD
code~\cite{CPMD} using the PBE generalized gradient
approximation~\cite{PBE} with $N$=64 atoms, a time step of $0.77$ fs,
Troullier-Martin pseudopotentials, and a 100 Ry cut-off for the plane
wave expansion of the Kohn Sham orbitals, combined with $\Gamma$ point
sampling of the Brillioun zone. A finite size study showed that the
Hugoniot results are well converged with $N$=64.


\begin{figure}[!]
\includegraphics[angle=0,width=\figurewidth]{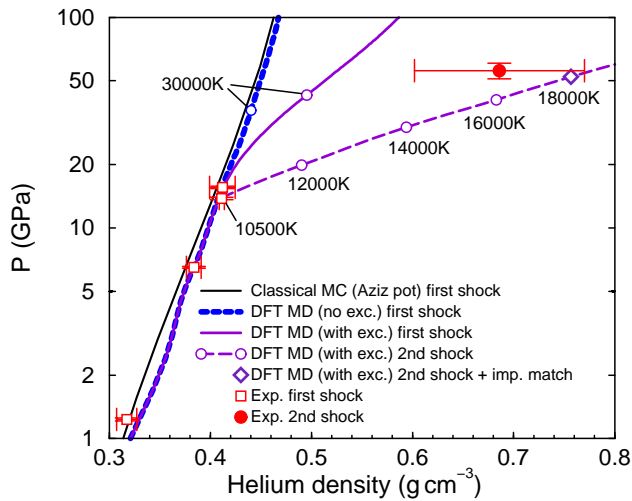}
\caption{ 
       The principal shock Hugoniot curves computed using DFT-MD with
       (solid line) and without (blue dashed line) the consideration
       of excited electronic states are compared with gas gun shock
       wave experiments~\cite{nellis84} (open red symbols). The dashed
       purple line shows the computed reshock curve. The $\diamond$
       on it indicates the impedance match condition for the
       reshock experiment~\cite{nellis84} (solid red symbol). }
\label{fig1}
\end{figure}

The only available shock data for fluid helium were obtained with gas
gun experiments by Nellis {\it et al.}~\cite{nellis84}. The comparison
shown in Fig.~\ref{fig1} shows excellent agreement between
experimental data and DFT-MD simulation results for the principal and
the reshock Hugoniot curves. 

\begin{figure}[!]
\includegraphics[angle=0,width=\figurewidth]{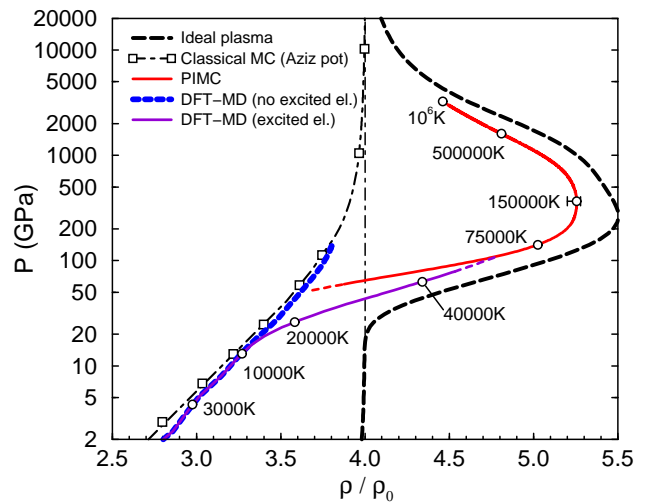}
\caption{ 
       The principal shock Hugoniot curve of helium is shown as a
       function of compression ratio. PIMC and DFT-MD results were
       combined to span a large range of temperature, as the circles
       in the curve indicate. Effects of thermal electronic
       excitations (purple solid line) were used to correct the
       ground state DFT-MD calculations (blue dashed line), which
       resulted in a strong increase in compressibility. Results
       a from classical simulation using the Aziz pair potential and
       from a non-interacting plasma model are included.}
\label{fig2}
\end{figure}

The DFT-MD results without electronic excitations are closely tracked
by data from classical Monte Carlo (CMC) simulations using the Aziz
pair potential~\cite{aziz95}. This potential was derived to describe
the interaction of two isolated helium atoms. With rising shock
pressure, DFT-MD and CMC curves in Fig.~\ref{fig2} show a gradual
increase in compression towards 4-fold the initial density, which
represents the high pressure limit, where the kinetic energy dominates
over the potential energy and the system behaves approximately like
non-interacting particles.

Conversely, the results from PIMC calculations predict much higher
compression ratios, reaching a maximum of 5.24$\pm$0.04 for $P=360$
GPa and $T=150\,000\,$K. This increase in compression is due to
electronic excitations that are present in the hot dense fluid at
thermodynamic equilibrium. Helium differs in this regard from shocked
deuterium, for which PIMC calculation predicted a maximum compression
ratio of only 4.3(1)~\cite{MC00}.

In our PIMC simulation program, the pair density matrix
technique~\cite{PC84,MG05} is employed to treat the Coulomb
interactions. The accurate treatment of the many-body correlations and
the application of the nodal restriction for the paths requires the
use of a small step, $\tau=\beta/M$, to discretize the paths in
imaginary time $\beta=1/k_BT$ into $M$ steps. Using a time step
between $\tau^{-1}=2\times 10^6$ and $16\times 10^6$K depending on
density, allowed a reduction of the remaining time step error in the
observables below the size of MC error bars.

PIMC simulations with $N_e=32$ electrons and $N$=16 nuclei were
performed on a grid of 12 densities and 5 temperatures ranging from $3.5
\ge r_s \ge 1.0$ and $10^6 \ge T \ge 61250\,$K. The resulting EOS was 
interpolated to obtain the Hugoniot curves. A finite-size
extrapolation with up to $N$=57 nuclei was done for four points in
$T$-$\rho$ space, $T$=125$\,$000 and 10$^6$$\,$K combined with
$r_s=$1.86 and 1.25. Above 100 GPa, the Hugoniot curve is insensitive
to finite size corrections because of the high temperatures and due to
the partial cancellation of pressure and internal energy corrections
in Eq.~\ref{hug}. After propagating the correction and the uncertainty
of the finite size extrapolation, we obtained 5.24$\pm$0.04 as maximum
compression ratio of fluid helium, which actually brackets the
original value of 5.25 obtained for $N$=16.

The predicted increase in compression beyond 4-fold the initial
density in helium can be understood by invoking a very simple free
energy model, which assumes helium to be composed of different
non-interacting species, He, He$^+$, and He$^{2+}$, to represent the
various ionization stages of helium as well as free electrons. The
resulting Hugoniot curve shows a similar increase in the compression
ratio demonstrating that the increase beyond 4-fold is due to
electronic excitations leading to free electrons.

To further verify this hypothesis, we corrected the DFT EOS for
finite-temperature electronic effects. For a number of snapshots along
the MD trajectory, we thermally populated the instantaneous excited
electronic states~\cite{Abinit} using the Mermin functional with up to
7 additional orbitals per atom. For temperatures above 15$\,$000$\,$K,
the resulting corrections to the internal energy and pressure leads to
a substantial increase in shock compressibility. This increase is
primarily caused by a rise in the internal energy due to thermal
population of excited electronic states. Fluid helium maintains a wide
excitation gap ranging between 5 and 15 eV for $T \le 80\,000\,$K and
$2.6 \ge r_s \ge 1.0$. On the principal Hugoniot, electronic
excitations occur above 20 GPa, which explains why the gas gun
experiments have not reached the regime of electronic
excitations. Even a gas gun reshock experiment would be insufficient
because the final temperatures remain relatively low
(Fig.~\ref{fig1}), and facilities that can generate faster shock waves
are needed instead.

The maximum shock pressures that can be reached at a particular
experimental facility depend on the power of the drive but also on the
impedance of the sample material. For the same shock drive, the final
shock pressure is, to a first approximation, proportional to the
initial density of the sample material. The maximum pressure reported
for deuterium Nova laser shocks~\cite{Si97,Co98}, 340 GPa, and the
highest achievable on the Z machine~\cite{Kn01,Kn03}, 175 GPa, would
translate to approximately 246 GPa, and 127 GPa~\footnote{A more
accurate estimate for Z machine including the impedance match yields
140 GPa. (M. Knudson, private communication.)} in
helium. Consequently, with both facilities one would be able reach the
regime of the predicted 5-fold compression and probe the effect of
electronic excitations.

\begin{figure}[!]
\includegraphics[angle=0,width=\figurewidth]{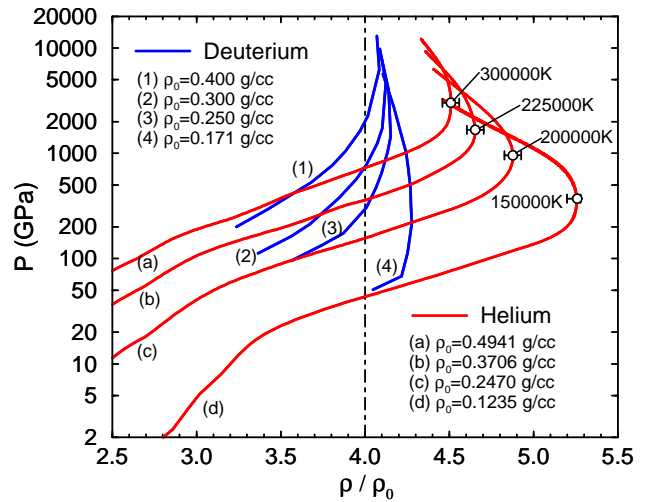}
\caption{ 
   The principal shock Hugoniot curves for deuterium~\cite{Mi03b} and
   helium are shown for samples that were precompressed to different
   initial densities. For both materials, the precompression reduces
   the maximum compression ratio $\rho/\rho_0$ that can be
   reached. For helium, the indicated initial densities (a)-(d)
   correspond to the initial pressures of 7.1 kbar, 1.8 kbar, 188
   bar~\cite{driessen86}, and 1 bar. The DFT-MD and PIMC data shown in
   Fig.~\ref{fig2} were interpolated for helium. }
\label{fig3}
\end{figure}

\begin{figure}[!]
\includegraphics[angle=0,width=\figurewidth]{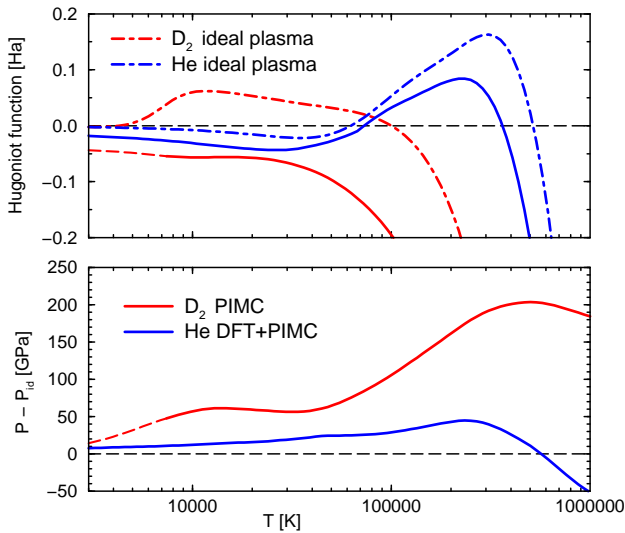}
\caption{ 
   The Hugoniot functions $H$ from Eq.~\ref{hug} for helium and
   deuterium derived from first principles calculation (solid lines)
   and from the non-interacting plasma model (dashed line) is shown at
   5-fold compression. Deuterium is less compressible than helium
   because the interaction between the particles is much stronger,
   which leads to a substantial increase in pressure beyond the
   corresponding value of the non-interacting system including
   ionization ($P_{\rm id}$), illustrated by the pressure difference
   in the lower graph.  }
\label{fig4}
\end{figure}

The comparison in Fig.~\ref{fig3} shows the discussed increase in
compressibility beyond 4-fold for helium, while our results for
deuterium~\cite{MC00,Mi01} show hardly any, despite the
similarity of the two fluids. Deuterium molecules and helium atoms
have the same mass and provide two mechanisms to absorb shock energy,
which in principle can lead to shock compression ratios substantially
larger than 4.  The helium atom has two ionization stages with
energies of 24.6 and 54.4 eV. Deuterium molecules dissociate with 4.5
eV energy, and the ionization of resulting atoms requires 13.6 eV.
However, the explanation for the different shock behavior of helium
and deuterium is not a consequence of single particle properties but
is a result of different degrees of particle interaction.
Fig.~\ref{fig4} shows the Hugoniot function, $H$, for both materials
at 5-fold compression. For helium, first principles calculation that
include the interaction effects, as well as the non-interacting plasma
model, predict that $H$ changes its sign, implying that helium is
more than 5-fold compressible. While the non-interacting plasma model
predicts the same behavior also for deuterium, the PIMC simulations
show that the $H$ function is strictly negative because the pressure
is substantially higher than suggested by the non-interacting plasma
model.

The interaction effects at 5-fold compression are stronger in
deuterium than in helium because in helium, the initial density is
lower and, more importantly, the electronic orbitals are much more
localized. The initial density, $\rho_0=0.1235\,$g$\,$cm$^{-3}$
corresponds to a Wigner-Seitz radius of $r_s=3.51$ [$\frac{4}{3}\pi
(r_s a_0)^3=V/N_e$], and 95\% of the electron charge is localized in
10\% of the total volume. The initial density of fluid deuterium,
$\rho_0= 0.171\,$g$\,$cm$^{-3}$ corresponds to $r_s=3.16$, and 95\% of
the charge is localized in 40\% of the volume. With increasing
compression, the orbitals in deuterium start to overlap sooner and the
resulting polarization and exchange interactions will increase the
pressure, which causes deuterium to appear less compressible in shock
experiments.


In Fig.~\ref{fig3}, we use our first principles calculations to
predict how the static precompression will affect the Hugoniot curves
of helium and deuterium~\cite{Mi03b}. For helium, a precompression to
4-fold the cryogenic liquid density at 1 bar, $\rho_{0}=4 \rho_{0}^{(1
{\rm bar})}$, increases the maximum density on the Hugoniot curve to
$\rho$ = 17.5 $\rho_{0}^{(1{\rm bar})}$. However, the maximum
compression ratio for the dynamic compression, $\eta=\rho/\rho_0$,
decreases in both materials with increasing precompression. The
increase in the initial density leads to stronger interactions under
shock conditions. In general the shock compression ratio is determined
by the relative importance of excitations of internal degrees of
freedom that increase the internal energy and interaction effects that
increase the pressure.

The point of maximum compression, $\eta_{\rm max}$, along the Hugoniot
curve is reached when the Gr{\"u}neisen parameter, $\gamma \equiv V
\left. \frac{\partial P}{\partial E}\right |_V$, satisfies $\gamma =
\frac{2}{\eta-1}$. A reduction in the compression ratio with
precompression can be expressed by $\frac{d \eta_{\rm
max}}{d V_0} > 0$. Using Eq.~\ref{hug} and assuming $E_0$ does
not change with the precompression, this condition can be expressed in
terms of the simple thermodynamic condition, $\frac{\rho}{P}
\left . \frac{\partial P}{\partial \rho}\right|_E > 1.$ We 
have computed the isoenergetic compressibility, $\left
. \frac{\partial P}{\partial \rho}\right|_E$, and verified that this
condition is satisfied for hydrogen and helium by using our first
principles results as well as with the non-interacting plasma
model. This confirms that the precompression reduces the maximum
compression ratio in hydrogen and helium, and one can postulate that
this might also be true for other simple fluids.

The computed EOS shows that helium and hydrogen behave differently at
high pressure, which has important implications for the interior
structure of solar and extrasolar giant planets where it is has been
predicted that the two fluid phases become immiscible at high
pressures~\cite{Stevenson77a}. Our calculations predict that helium is
5.24(4)-fold compressible under shock conditions, which distinguishes
it from fluid deuterium, for which first principles calculations and
recent experiments~\cite{Kn01,Kn03,Belov2002,BoriskovNellis05}
predicted a maximum compression ratio close to 4.3. We suggest that
all deuterium experiments be repeated with helium in order to validate
the different shock compression techniques. This validation is
important before the EOS can be used for wide range of applications
including the modeling of giant planets and to draw conclusions about
their evolution.

The author acknowledges fruitful discussions with N. Ashcroft,
J.D. Johnson, R. Hemley, R. Cohen, E.L. Pollock, S. Gramsch,
I. Tamblyn, and J. Vorberger. This material is based upon work
supported by NASA under the grant NNG05GH29G and by the NSF under the
grant 0507321.
%
%

\end{document}